# Machine learning analysis of tunnel magnetoresistance of magnetic tunnel junctions with disordered MgAl$_2$O$_4$

Shenghong Ju[1,2,3], Yoshio Miura[3,4,5,6], Kaoru Yamamoto[4,6], Keisuke Masuda[4,6], Ken-ichi Uchida[2,3,4,6,7,8], and Junichiro Shiomi[2,3,6]

[1] China-UK Low Carbon College, Shanghai Jiao Tong University, No. 3 Yinlian Road, Lingang, Shanghai 201306, China

[2] Department of Mechanical Engineering, The University of Tokyo, 7-3-1 Hongo, Tokyo 113-8656, Japan

[3] Center for Materials Research by Information Integration (CMI2), Research and Services Division of Materials Data and Integrated System (MaDIS), National Institute for Materials Science (NIMS), 1-2-1 Sengen, Tsukuba 305-0047, Japan

[4] Research Center for Magnetic and Spintronic Materials, National Institute for Materials Science (NIMS), 1-2-1 Sengen, Tsukuba 305-0047, Japan

[5] Center for Spintronics Research Network (CSRN), Graduate School of Engineering Science, Osaka University, Machikaneyama 1-3, Toyonaka, Osaka 560-8531, Japan

[6] Core Research for Evolutional Science and Technology (CREST), Japan Science and Technology Agency (JST), 4-1-8, Kawaguchi, Saitama 332-0012, Japan

[7] Institute for Materials Research, Tohoku University, Sendai 980-8577, Japan

[8] Center for Spintronics Research Network, Tohoku University, Sendai 980-8577, Japan

**ABSTRACT**

Through Bayesian optimization and the least absolute shrinkage and selection operator (LASSO) technique combined with first-principles calculations, we investigated the tunnel magnetoresistance (TMR) effect of Fe/disordered-MgAl$_2$O$_4$(MAO)/Fe(001) magnetic tunnel junctions (MTJs) to determine structures of disordered-MAO that give large TMR ratios. The optimal structure with the largest TMR ratio was obtained by Bayesian optimization with 1728 structural candidates, where the convergence was reached within 300 structure calculations. Characterization of the obtained structures suggested that the in-plane distance between two Al atoms plays an important role in determining the TMR ratio. Since the Al-Al distance of disordered MAO significantly affects the imaginary part of complex band structures, the majority-spin conductance of the $\Delta_1$ state in Fe/disordered-MAO/Fe MTJs increases with increasing in-plane Al-Al distance, leading to larger TMR ratios. Furthermore, we found that the TMR ratio tended to be large when the ratio of the number of Al, Mg, and vacancies in the [001] plane was 2:1:1, indicating that the control of Al atomic positions is essential to enhancing the TMR ratio in MTJs with disordered MAO. The present work reveals the effectiveness and advantage of material informatics combined with first-principles transport calculations in designing high-performance spintronic devices based on MTJs.

**Keywords:** Tunnel magneto-resistance, magnetic tunnel junction, disordered spinel barrier, density functional theory, Bayesian optimization

## I. INTRODUCTION

Magnetic tunnel junctions (MTJs), which exhibit different resistance depending on the relative magnetization directions of the ferromagnetic electrodes, are among the most important devices for spintronic applications, such as magnetic random access memories (MRAMs) and high-sensitivity magnetic sensors [1,2]. MgO-based MTJs have yielded large tunnel magnetoresistance (TMR) ratios of over 600% at room temperature owing to coherent tunneling through the $\Delta_1$ evanescent state of MgO and the half-metallic behavior of the $\Delta_1$ state in Fe and CoFe [3-6]. However, epitaxial growth of crystalline MgO is difficult because of its lattice mismatch with the ferromagnetic electrode, which is about 4% for bcc-Fe and over 5% for half-metallic Co-based full Heusler alloys. Recently, Sukegawa and coworkers have fabricated new epitaxial barrier layers, $MgAl_2O_4$ on Fe electrodes, and initially observed a TMR ratio of about 165% at 15 K [7]. Normal spinel-type $MgAl_2O_4$ (MAO) has two different cation sites: tetragonal A-sites and octahedral B-sites. The lattice constant of MAO is about 8.09 Å, which leads to a very small lattice mismatch of less than 0.5%, with the (001) face of bcc Fe having a 45 degree (001) in-plane rotation. Furthermore, the lattice constant of the spinel barrier can be modulated by choosing various metallic elements for the cation atoms in A-site and B-site, making it a promising candidate for the new barrier layer.

Density functional theory (DFT) calculations yielded a TMR ratio of 160% for Fe/MAO/Fe MTJs [8], which is one order of magnitude smaller than the 1600% obtained for Fe/MgO/Fe MTJs [3,4]. Since the in-plane lattice constant of MAO-based

MTJs is twice that of MgO-based MTJs, boundary edge states in the two-dimensional Brillouin zone are folded to the Γ point ($k_x=k_y=0$). The band folding effect opens a new conductive channel in the minority-spin states at the Γ point, reducing the spin polarization of the tunneling conductance and the TMR ratio in Fe/MAO/Fe MTJs.

Earlier experimental studies have reported an observed TMR ratio for MAO-based MTJs of about 165% at 15 K [7], which was consistent with the calculated results [8]. Then, higher TMR ratios of 328% and 479% were observed at 15 K for Fe/MAO/Fe and CoFe/MAO/CoFe MTJs, respectively. This enhancement in TMR ratio resulted from the suppression of the band-folding effect of disordered MAO. In MAO-based MTJs showing relatively low TMR ratios (165%), the barrier layer was fabricated by oxidizing the Al layer on Mg, and the lattice constant estimated from the diffraction pattern was about 8 Å, which corresponds to the lattice constant of bulk MAO. On the other hand, MAO-based MTJs showing high TMR ratios of over 300% at 15 K were fabricated through oxidation of the Mg-Al layer. In this case, the lattice constant of the barrier layer was 4.0 Å, which corresponds to half that of bulk MAO. Sukegawa *et al.* found that the unit cell size is reduced when the cations $Mg^{2+}$ and $Al^{3+}$ in spinel MAO randomly occupy octahedral and tetrahedral sites including vacant sites [9]. Cation site disordering changes the space group to either F-m3m or F-43m and halves the unit cell size of MAO, while the anion sites are always occupied by oxygen atoms, and the fcc sublattice of oxygen is not changed by the disordering.

Cation-site-disordered MAO yields higher TMR ratios than ordered MAO owing to suppression of the band folding effect. However, the TMR ratios observed were still lower than those of MgO-based MTJs. In order to enhance the TMR ratio of MAO-based MTJs, it is important to identify the local structures of cation-site-disordered MAO, which leads to high TMR ratios. However, at the present time, the detailed local structure of cation-site-disordered MAO cannot be determined experimentally. Furthermore, determining the optimal structure through the DFT method alone is challenging, because the number of atomic configurations of disordered MAO increases sharply with increasing MTJ unit cell size.

Materials informatics (MI), which combines data science and traditional simulation/experiments, has become a popular tool to accelerate the materials discovery and design process [10,11]. For instance, in the development of thermal functional materials, MI has greatly reduced the computational cost, and the optimal nanostructures for phonon transport could be calculated by using only a few percent of the total candidate structures through machine learning methods such as Bayesian optimization [12] and Monte Carlo tree search [13]. Various regression models including the Gaussian process and support vector regression models have been applied to predict the thermal boundary resistance, and the results indicate that machine learning models have much better predictive accuracy than the commonly used acoustic mismatch model and diffuse mismatch model [14,15]. In the case of thermoelectric materials, Bayesian optimization has been used improve the thermal and electronic properties simultaneously. As a result, the thermoelectric performance of defective graphene nanoribbons was enhanced by 11-times compared with that of pristine ones [16]. Hou *et al.* [17] also adopted Bayesian optimization to optimize the Al/Si ratio in Al$_2$Fe$_3$Si$_3$ and improved the power factor by 40%, which significantly reduced the time and labor cost of optimizing thermoelectric materials. More recently, Bayesian optimization has been used to design wavelength-selective thermal radiators [18], which can serve as thermophotovoltaics and infrared heaters. On the basis of these studies, we concluded that complementing DFT calculations with Bayesian optimization would allow us to determine the optimal disordered MAO structure. Thus, in the present work, we combine MTJ calculations with Bayesian optimization to design a disordered spinel barrier structure, aiming to maximize the TMR ratio and elucidate the underlying mechanism.

## II. METHODOLOGY

We constructed the supercell of a Fe/disordered-MAO/Fe (001) MTJ containing 10 atomic layers of bcc Fe and 3 atomic layers of disordered MAO, as shown in Fig. 1. The rock-salt structure was assumed for disordered MAO, where the cation sites are randomly occupied by Al, Mg, <u>and</u> a vacancy. This kind of defective rock-salt structure has already been experimentally demonstrated [9]. We prepared two types of supercells with different in-plane unit cell sizes for the Fe/MAO/Fe MTJ. One had cross-sectional size of 5.732 Å × 5.732 Å, which is the smallest in-plane unit cell of ordered MAO. The other was 5.732 $\sqrt{2}$ Å × 5.732 $\sqrt{2}$ Å in cross-section, which corresponds to a 45° (001) in-plane rotation of the former structure. 5.732 Å is twice the lattice constant of bcc Fe (2.86 Å).

The TMR ratio for each MTJ structure is defined as,

$$\text{TMR ratio} = \frac{R_{\text{AP}} - R_{\text{P}}}{R_{\text{P}}}, \tag{1}$$

where $R_{\text{AP(P)}}$ is the electrical resistance in the anti-parallel (parallel) state. The TMR ratio was obtained by first-principles calculations within the generalized-gradient approximation for exchange-correlation energy by using the Atomistix ToolKit simulation package (ATK) [19]. In the transport calculations, which were based on the Landauer formula, ballistic tunneling conductance of the infinite open system of Fe/disordered-MAO/Fe MTJs attached to 4 atomic layers of left and right Fe electrodes was obtained by using the atomic Green function in ATK. The number of k-points was taken to be 7×7×50 for self-consistent DFT calculations, and 100×100 for transport calculations.

By assigning the Al, Mg, and vacancy positions to the rock-salt structure sites, various candidate structures can be obtained. To find the global optimal structure with the highest TMR ratio, we adopted Bayesian optimization, which is effective in determining the optimal candidate with high efficiency. The framework of the MI method, which combines MTJ calculations and Bayesian optimization, is illustrated in Fig. 2. First, we randomly selected 20 structures for initial accurate TMR ratio calculations. Then, a Bayesian regression model was trained by the 20 pairs of descriptors and TMR ratios that were collected. On the basis of this predictive model, the transport properties of the remaining candidates were evaluated. The top 20 candidates with the highest estimated TMR ratios are then selected for the next round of accurate calculations. The newly obtained accurate TMR ratio is added to the training examples to update and improve the model. This procedure is repeated, and the final

TMR ratio to be obtained is taken as the global optimal. The best disordered spinel barrier structure can thus be quickly determined.

## III. RESULTS AND DISCUSSION

### A. Performance of Bayesian optimization

First, the performance of Bayesian optimization was tested for a system with smaller cross-section. We fixed the number of vacancy sites in each layer of the rock-salt structure so that the stoichiometry of Mg:Al:O was maintained at 1:2:4. According to combination theory, this gives a total of 1728 candidate structures. In Bayesian optimization, we set the descriptor to identify atomic structures of disordered MAO. The adopted descriptors were integer flag values to describe the atoms occupying the sites of the rock-salt structure, where "2", "1", and "0" represented the Mg, Al, and vacancy, respectively. For comparison, 10 rounds of Bayesian optimization and random search were conducted with different choices of initial candidates, as shown in Figs. 3(a) and 3(b). All Bayesian optimizations reached convergence within 300 structure calculations, while random search required at most around 1500 calculations. Figure 3(c) shows the averaged trends of convergence, which clearly indicate that Bayesian optimization is much more efficient than random search. The total number of calculations needed to obtain the global optimal structure is 17%, which is somewhat larger than in phonon transport optimization [11]. This is mainly due to that the total number of candidates in the current case is relatively small, as the efficiency of machine learning is usually limited for small data. Therefore, the efficiency is expected to increase with increasing number of candidates.

To check the accuracy of Bayesian optimization, we performed a full calculation for all of the 1728 candidates. The global maximum TMR ratio reached as high as 600.18%, and the corresponding structure was confirmed to be exactly the same as that obtained by Bayesian optimization. Figure 3(d) shows the TMR ratio distribution. There are few candidate structures located in the high-TMR-ratio region, which indicates that Bayesian optimization is suitable to search for MTJs with the highest TMR ratio.

Figure 4 shows the in-plane wave-vector dependence of the tunneling conductance at the Fermi level for the Fe/MAO/Fe MTJ showing the highest TMR ratio of 600% in parallel and antiparallel magnetization configurations. The majority-spin conductance for the parallel magnetization shown in Fig. 4(a) has a broad peak around the center of the 2D Brillouin zone, which is much stronger than the minority-spin conductance and the conductance for antiparallel magnetization. This agrees with the typical behavior of the coherent tunneling conductance of $\Delta_1$ electrons at the Brillouin zone center. Furthermore, the in-plane wave-vector dependence in antiparallel magnetization has no peak around $(k_x, k_y)=(0,0)$ and shows a hot-spot-like peaked structure in the 2D Brillouin zone. This means that the contribution of the $\Delta_1$ evanescent state to the tunneling conductance is completely blocked in the MTJ because of the suppression of the band holding effect in disordered MAO.

**B. Effect of cation site distance on TMR ratio**

To identify local structures of MAO with high TMR ratios, we performed the Least Absolute Shrinkage and Selection Operator (LASSO) regression of the TMR ratio for Fe/disordered MAO/Fe with 1728 cases. As a descriptor for the LASSO regression, we considered the averaged atomic distance between all the cation atoms of the MAO including vacancies (Vc). We considered two kinds of atomic distances: (i) the averaged distance along the in-plane direction of the MTJs (dXY), and (ii) the averaged distance along the normal to the (001) plane (dZ). dXY and dZ are physically different because of the non-periodicity of the MTJ along the normal to the (001) plane. Figure 5(a) shows the coefficients of LASSO regression for each descriptor as a function of the LASSO regularization strength $\alpha$. The increase of $\alpha$ forces all of the regression coefficients to zero. As can be seen in Fig. 5(a), the descriptors of dXY(Al-Al) and dXY(Al-Vc) have very large coefficients as compared with other descriptors. Furthermore, the coefficient of dXY(Al-Al) survives for larger $\alpha$. This means that the in-plane distance of Al-Al is essential for characterizing the TMR of the Fe/disordered-MAO/Fe system. In Fig. 5(b), we compare the TMR ratio obtained from first-principles calculations and the TMR ratio estimated by LASSO with $\alpha = 0.1$. These results show that the TMR ratios predicted by LASSO qualitatively agree with those obtained explicitly by using first-principles calculations. Since the TMR ratios of Fe/disordered-MAO/Fe MTJs are very sensitive not only to the cation-site atomic distances but also interfacial structures, descriptors related to interfacial structures will be required to obtain quantitative agreement between LASSO results and first-principles results.

Figures 6(a) and 6(b) show the correlation between TMR ratios from first-principles calculations and in-plane distances (dXY) of Al-Al and Al-Vc. As seen in Figs 5(a) and (b), the TMR ratio tends to increase with increasing dXY(Al-Al) and decrease with increasing dXY(Al-Vc). Since there are four cation sites in the [001] plane of disordered MAO, with cross-sectional size of 5.732×5.732 Å, two Al atoms tend to be located in the diagonal positions for larger dXY and they tend to be located in the vertical or horizontal positions for smaller dXY (see Fig. 1(a)). Thus, we can say that the TMR ratio tends to increase when two Al atoms are located in the diagonal positions. To clarify this point, Figs. 6(c) and 6(d) show the correlation between TMR ratio and tunneling conductance in the parallel magnetization configuration. We found that there is a correlation between TMR ratio and majority-spin conductance: TMR ratio increases with increasing majority-spin conductance in parallel magnetization, as expected on the basis of the definition of the TMR ratio. By contrast, the correlation between TMR ratio and minority-spin conductance is very weak. This means that the high TMR ratio in Fe/disordered-MAO/Fe MTJs is due to the large majority-spin conductance in the parallel magnetization configuration.

Figures 7(a) shows the correlation between dXY(Al-Al) and majority-spin conductance in the parallel magnetization configuration. It can be seen that the majority-spin conductance tends to increase with increasing dXY(Al-Al). That is, the majority-spin conductance increases when two Al atoms are located in the diagonal positions (larger dXY). Since the tunneling conductance of Fe/disordered-MAO/Fe(001) MTJs can be characterized by the $\Delta 1$ evanescent in the barrier layer, it would be useful to check the complex band structure of disordered MAO in the bulk configuration. The imaginary part of the complex band structure κ gives the decay rate of the evanescent wavefunction in the barrier layer in the form exp(-κd), where $d$ is the thickness of the barrier layer. Thus, the state with the smallest $\kappa$ mainly contributes to the tunneling conductance. We calculated the imaginary part of the complex band structure for disordered MAO with 1728 cases in the bulk configuration. Fig. 7(b) shows the correlation between the in-plane Al-Al distance dXY(Al-Al) and the smallest $\kappa$ of bulk disordered MAO. We can find that the smallest $\kappa$ decreases with increasing dXY(Al-Al), leading to the larger majority-spin conductance and TMR ratio for larger dXY(Al-Al). Since the valence band of MAO is mainly composed of the Al 2$p$ orbitals, the in-plane Al-Al distance affects the complex band structure of MAO. As dXY(Al-Al) increases, the hybridization of the Al-Al 2$p$ wavefunction decreases, reducing the smallest imaginary part of the complex band for bulk disordered MAO. All these results indicate that disordered MAO, with its relatively large in-plane Al-Al distance, significantly affects the complex band structure of MAO and leads to the larger majority-spin conductance and TMR ratios of Fe/disordered MAO/Fe(001) MTJs. Thus, controlling the Al atomic position is key to obtaining larger TMR ratios in Fe/disordered MAO/Fe(001) MTJs.

## C. Effect of concentration distribution on TMR ratio

Having demonstrated the effectiveness of Bayesian optimization for MTJ design, we turn our attention to the MTJ with larger cross-section shown in Fig. 1(b). Instead of using a constant Mg:Al:vacancy distribution in the disordered spinel barrier region, we tried the three types of distributions listed in Table I. For each distribution, we performed Bayesian optimization separately, and the highest TMR ratios obtained were 153.8%, 312.5%, and 373.2%. We found that the maximum TMR ratios of the MTJs in Table I were not the same as those of the MTJs with a unit cell size of ($a \times a$) shown in Fig. 3. Since a 45° rotation changes the periodicity of the in-plane directions, we cannot obtain the same atomic configuration including the u-parameter of oxygen atoms for disordered MAO with a unit cell of ($\sqrt{2}a \times \sqrt{2}a$) and ($a \times a$). However, we confirmed that the physical cause of the maximum TMR ratio in MTJs with ($\sqrt{2}a \times \sqrt{2}a$) unit cell is almost the same as that of MTJs with ($a \times a$) unit cell. The results are summarized in Fig. 8, which reveals that a higher TMR ratio can be achieved with a more uniform Mg:Al:vacancy ratio distribution in the disordered spinel barrier region. This is useful experimental information that can aid the design of future MTJs. The present results indicate that in order to obtain a large TMR ratio using MTJs with disordered MAO, it would be better to have the Al atoms in the MAO barrier distributed uniformly. This is because such a uniform distribution of Al atoms would tend to increase the in-plane Al-Al distances. On the other hand, if the distribution of Al atoms is fluctuated, the averaged Al-Al distance in the same plane will decrease. This would increase the imaginary part of the complex band structure for disordered MAO, as discussed in the previous section, leading to a decrease in the majority-spin conductance and TMR ratio of Fe/disordered MAO/Fe MTJs.

## IV. CONCLUSION

We have optimized the disordered spinel barrier in MTJs to obtain a high TMR ratio using Bayesian optimization. The TMR ratio of Fe/disordered MAO/Fe MTJs is successfully optimized by Bayesian optimization. The maximum TMR ratio obtained is over 600%, which is much larger than the TMR ratio of Fe/ordered MAO/Fe MTJs (160%). We found that the in-plane distance between two Al atoms plays an important role in determining the TMR ratio from LASSO analysis. Since the increase in the Al-Al distance of disordered MAO reduces the imaginary part of the $\Delta_1$ evanescent state in complex band structures of disordered MAO, the majority-spin conductance of the $\Delta_1$ state in Fe/disordered-MAO/Fe MTJs increases with increasing in-plane Al-Al distance, leading to larger TMR ratios. Furthermore, our optimization results <u>reveal</u> that a uniform Mg:Al:vacancy ratio distribution in the disordered spinel barrier region is preferable for achieving a high TMR ratio. All these results indicate that controlling the Al atomic position is crucial for obtaining a large TMR ratio in Fe/disordered MAO/Fe(001) MTJs.

**ACKNOWLEDGMENTS**

The authors are grateful to H. Sukegawa and S. Mitani for useful discussions and helpful comments. The calculations in this work were performed using the supercomputer facilities of the Institute for Solid State Physics at the University of Tokyo. This work was supported in part by the "Materials Research by Information Integration" Initiative (MI2I) Project, CREST "Scientific Innovation for Energy Harvesting Technology" (Grant No. JPMJCR16Q5) and CREST "Creation of Innovative Core Technologies for Nano-enabled Thermal Management" (Grant No.

JPMJCR17I1) from the Japan Science and Technology Agency (JST) and KAKENHI Grant-in-Aid for Early-Career Scientists (Grant No. 19k14902) from the Japan Society for the Promotion of Science (JSPS).

**Figures, tables and captions**

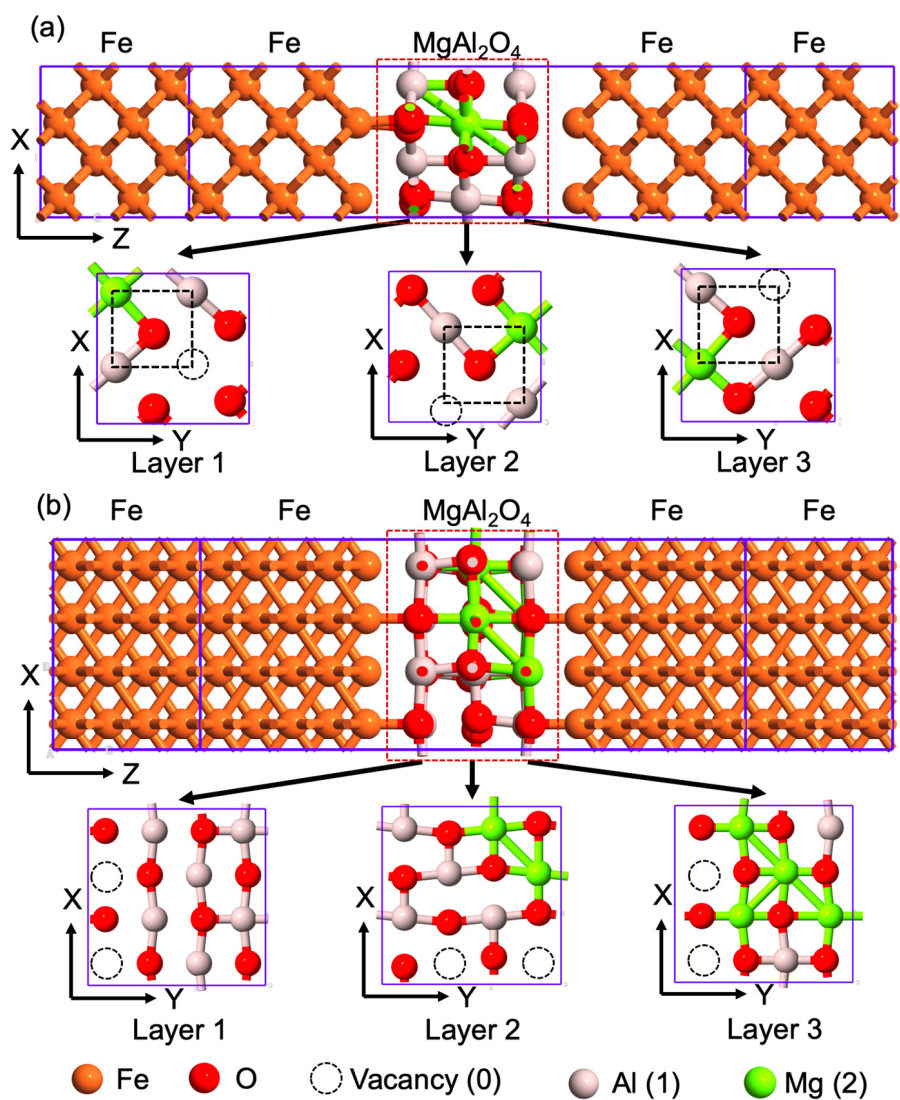

**Fig. 1** (a) System for Fe/MAO/Fe optimization with cross-sectional size of 5.732 Å × 5.732 Å. The crystal structure of disordered MAO was fixed to a rock-salt type structure with 3 atomic layers, and 1728 candidate structures were considered. (b) Fe/MAO/Fe system with 5.732 $\sqrt{2}$ Å × 5.732 $\sqrt{2}$ Å cross-section.

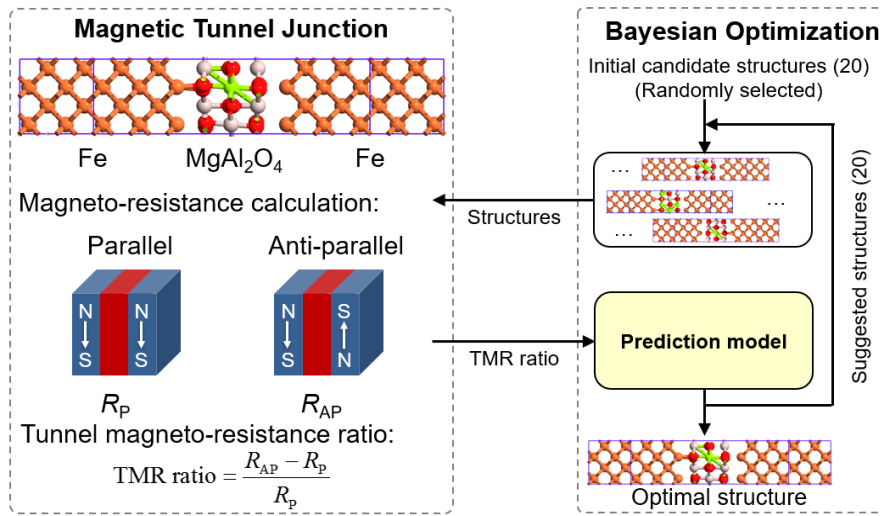

**Fig. 2** Schematics of designing magnetic tunnel junction for high tunnel magneto-resistance ratio by Bayesian optimization.

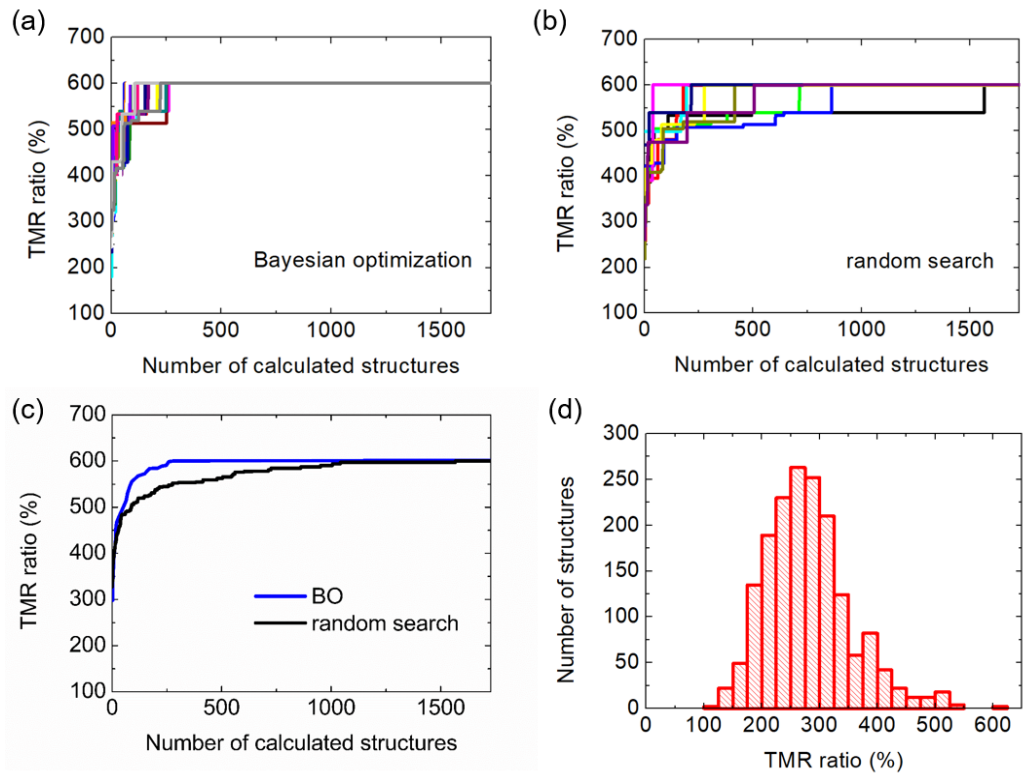

**Fig. 3** Performance of Bayesian optimization for a Fe/MAO/Fe system with 5.732 Å × 5.732 Å cross-section. (a) and (b) show 10 optimization runs each, for Bayesian optimization and random search, respectively. (c) Comparison of Bayesian optimization with random search. (d) TMR ratio histogram.

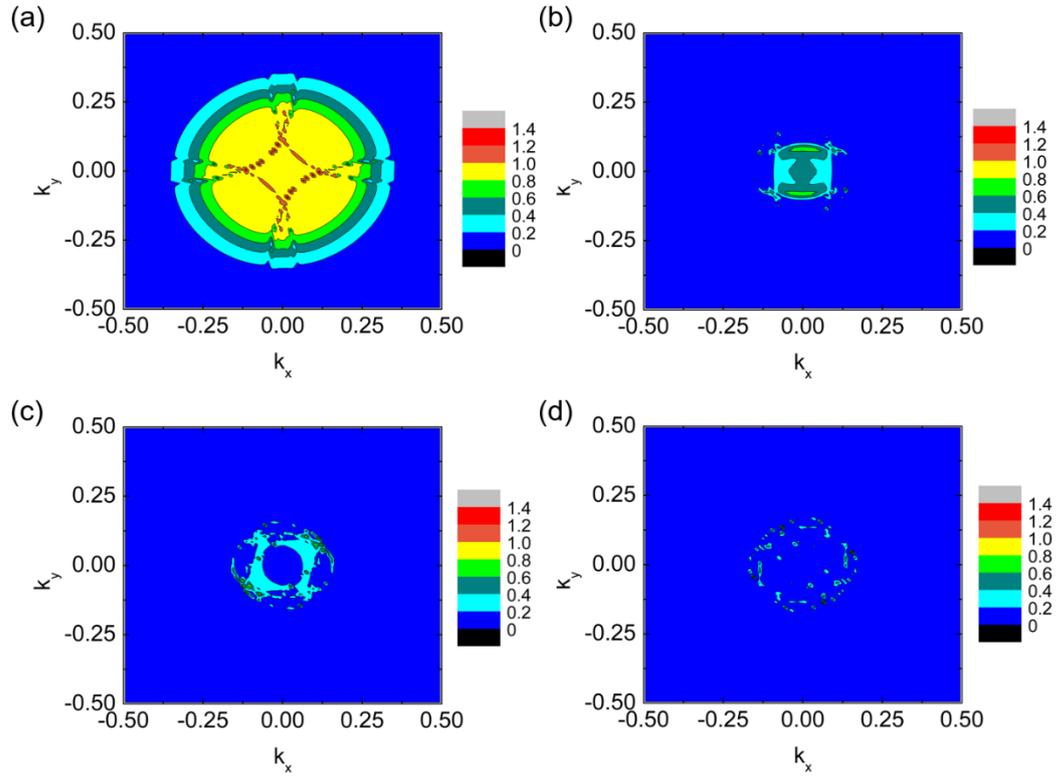

**Fig. 4** In-plane wave-vector dependence of the transmission at the Fermi energy for a Fe/MAO/Fe MTJ with 5.732 × 5.732 Å cross-section. (a) majority-spin, parallel, (b) minority-spin, parallel, (c) majority-spin, antiparallel, (d) minority-spin, antiparallel.

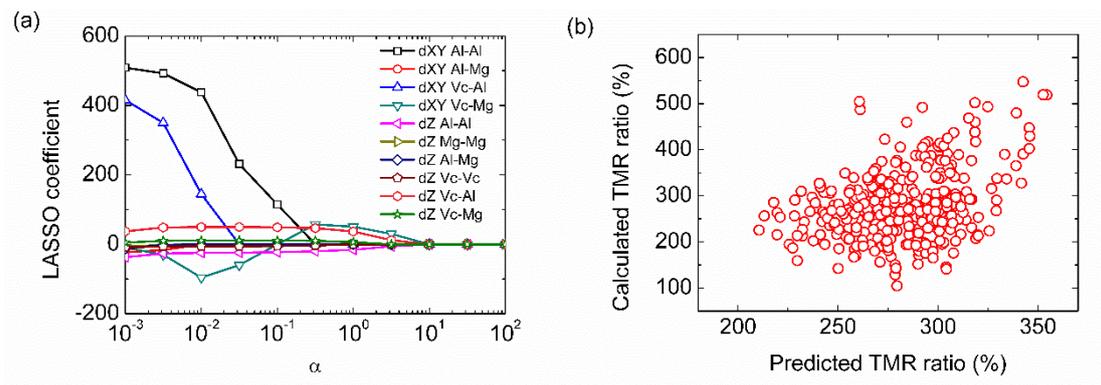

**Fig. 5** Coefficients of LASSO regression for TMR ratio by distance of each atomic site. dXY denotes the in-plane atomic distance, and dZ denotes the atomic distance perpendicular to the plane.

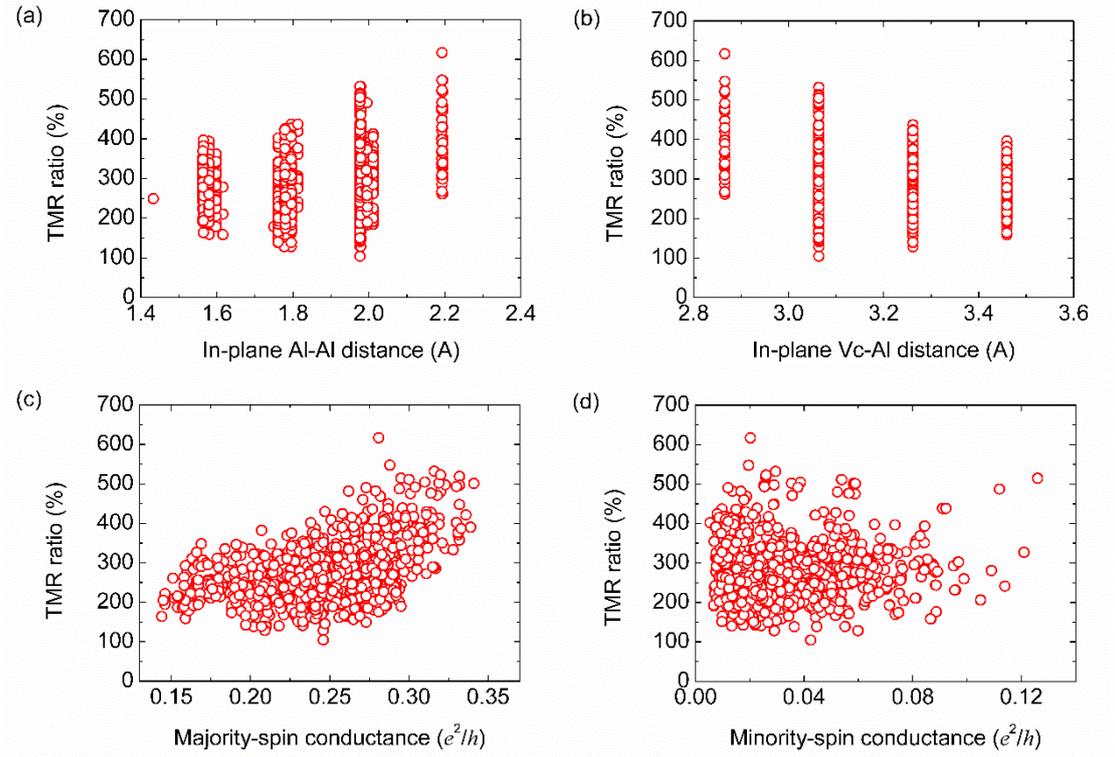

**Fig. 6** Correlation between TMR ratio and (a) in-plane Al-Al distance, (b) in-plane Vc-Al distance. The maximum TMR ratio increases (decreases) with increasing Al-Al (Vc-Al) distance. Vc denotes cation vacancies in the rock-salt structure. Correlation between TMR ratio and (c) majority-spin conductance and (d) minority-spin conductance in parallel magnetization.

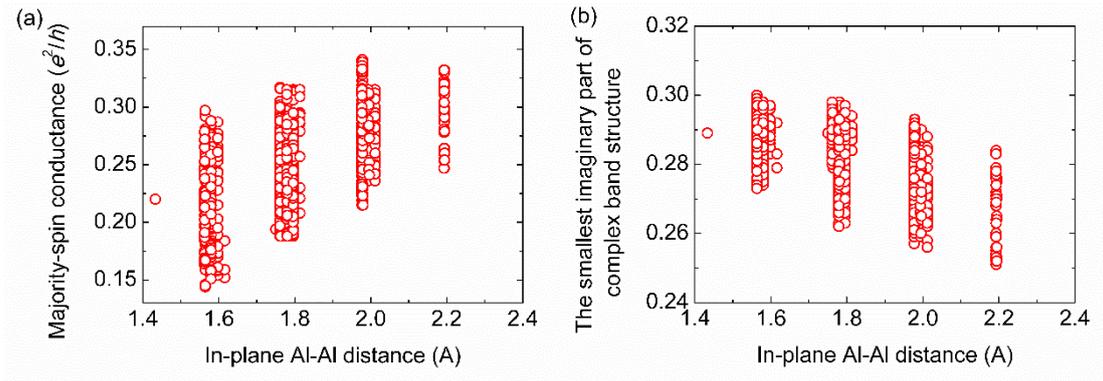

**Fig. 7** Correlation between (a) majority-spin conductance in parallel magnetization ratio and in-plane Al-Al distance, (b) smallest imaginary part of complex band structure of disordered MAO, $\kappa$, and in-plane Al-Al distance. Majority-spin conductance increases with increasing in-plane Al-Al distance, because the decay rate $\kappa$ in disordered MAO decreases with increasing in-plane Al-Al distance.

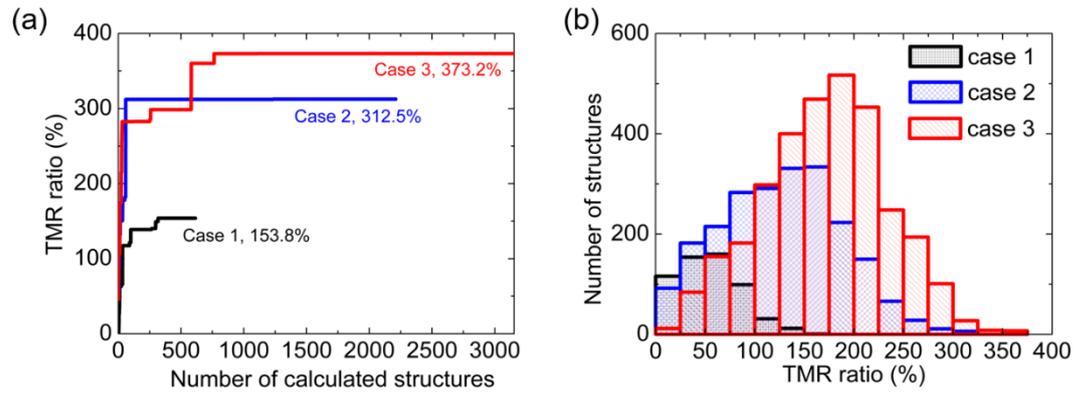

**Fig. 8** Optimization of Fe/MAO/Fe MTJs with 5.732 $\sqrt{2}$ Å × 5.732 $\sqrt{2}$ Å cross-sections. (a) Convergence of three cases for different Mg:Al:vacancy distributions in the disordered spinel barrier region. (b) TMR ratio histogram shows that case 3 with uniform Mg:Al:vacancy distribution has the highest TMR ratio.

**Table 1.** Optimized Fe/MAO/Fe MTJs with $5.732\sqrt{2} \times 5.732\sqrt{2}$ Å cross-sections.

| Case | Calculated / Total number of candidates | Mg:Al:vacancy ratio | | | Optimal TMR ratio |
|---|---|---|---|---|---|
| | | 1st layer | 2nd layer | 3rd layer | |
| 1 | 600/11,760 | 0:6:2 | 2:4:2 | 4:2:2 | 153.8% |
| 2 | 2,212/94,080 | 1:5:2 | 2:4:2 | 3:3:2 | 312.5% |
| 3 | 3,121/176,400 | 2:4:2 | 2:4:2 | 2:4:2 | 373.2% |